\documentclass[aps,pra,showpacs,floatfix,twocolumn,letterpaper,amsmath,amssymb]{revtex4}
\usepackage[pdftex]{graphicx}
\usepackage{amsmath}
\usepackage{hyperref}
\usepackage{latexsym}
\usepackage{color}
\usepackage{amssymb}
\usepackage{amsmath}
\usepackage{bm}

\newcommand{\mtrx}[1]{\ensuremath{\bm{\mathsf{#1}}}}

\usepackage{graphicx}


\begin{document}
\title{Molecular Hydrogen: a benchmark system for Near Threshold Resonances at higher partial waves}

\author{D. Shu, I. Simbotin, R. C\^ot\'e}
\affiliation{University of Connecticut,Storrs,CT 06268, USA}


\begin{abstract}
Benchmark reactions involving molecular hydrogen, such as H$_2$+D or H$_2$+Cl, provide
the ideal platforms to investigate the effect of Near Threshold Resonances (NTR) on
scattering processes. Due to the small reduced mass of those systems, shape resonances
due to particular partial waves can provide features at scattering energies up to a few Kelvins,
reachable in recent experiments. We explore the effect of NTRs on elastic and inelastic 
scattering for higher partial waves $\ell$ in the case of H$_2$+Cl for $s$-wave and H$_2$+D
for $p$-wave scattering, and find that NTRs lead to a different energy scaling of
the cross sections as compared to the well known Wigner threshold regime. We give
a theoretical analysis based on Jost functions for short range interaction potentials. To
explore higher partial waves, we adopt a three channel model that incorporates all
key ingredients, and explore how the NTR scaling is affected by $\ell$. A short discussion on 
the effect of the long-range form of the interaction potential is also provided.
\end{abstract}
\maketitle

\section{introduction}

Atomic and molecular hydrogen are the most abundant, and in many
ways the most fundamental, constituents of matter in the universe. For
example, reactions such as H$_2$+D$\rightarrow$HD+H are
relevant to astrophysics, especially for the astrochemistry in the
early universe \cite{astro} and the evolution of cold molecular
clouds in the earliest stages of star formation \cite{astro-cloud}.
In addition, hydrogen
is the perfect system to test our theoretical understanding by allowing
high precision calculations. In particular, basic chemical systems involving
molecular hydrogen, such as H$_2$+D, H$_2$+Cl, or H$_2$+F, provide
benchmark systems for which potential energy surfaces (PES) can be
calculated to a high level of accuracy.

Another fundamental feature in scattering are resonances. They are ubiquitous,
appearing as potential or shape resonance, or as Feshbach resonances. Although 
their effect is often averaged over at room or higher temperatures, they can become the 
dominant feature at low or ultralow temperatures, where only a few partial waves 
might contribute to the scattering process. Since cold molecules were
first predicted \cite{cote-1997,jmp-1999} and observed experimentally
\cite{pillet-1998,knize-1998}, rapid progress has been made in our ability to form and 
manipulate ultracold molecules \cite{Carr-NJP-2009,Dulieu-Review-2011}, which provides the 
seed to study in a precise and controlled fashion \cite{Quemener-review} the role of single 
partial waves, and state-to-state processes \cite{Chandler-2010} in chemical systems.
In fact, early experiments on KRb ultracold molecules \cite{Ospelkaus2010,deMiranda2011}, 
which explored quantum-state controlled chemical reactions using quantum statistics, 
motivated several studies of chemical systems under extreme conditions, and particularly 
the role of resonances in controlling the outcome. 
This high level of control over interactions can be realized using Feshbach 
resonances \cite{FR-RMP-2010}, or by orienting ultracold molecules 
\cite{paper-JILA,Jason-PRL}. In addition to investigations of
degenerate quantum gases \cite{RMP-bose-2001,RMP-fermi-2008}, such control also 
allows studies of exotic three-body Efimov states \cite{efimov,efimov-NTR} or application to 
quantum information processing \cite{DeMille-QC,Lena-1,Lena-2}.

In previous work, we have studied ultracold reactions involving molecular hydrogen,
such as H$_2$+D \cite{PCCP-H2+D,Rydberg-tuning,NJP-para-ortho-H2+D} 
H$_2$+Cl \cite{NTR-PRA,Simbotin2015} and H$_2$+F \cite{Simbotin2015}. We
also showed that shape resonances due to higher partial waves appear in 
certain reaction channels in H$_2$+D \cite{Rydberg-tuning,NJP-para-ortho-H2+D} 
and D$_2$+H \cite{D2+H-arXiv}. Due to the small reduced mass of these systems
and shallow van der Waals complexes, these resonances occur at scattering energies 
corresponding to a few mK up to a few K, {\it i.e.} above the ``standard"
ultracold regime usually nearing $\mu$K.
Similar resonances were recently observed in H$_2$ scattering with He$^*$ 
(metastable helium)  \cite{Narevicius-2012-Science,Narevicius-2014-NatChem,Narevicius-2015-NatChem,Narevicius-2016-NatPhys}. Such studies of benchmark
reactions involving H$_2$ should lead to a better understanding of the energy surface
and of the relevant scattering processes.

Previously, we explored the effect of near-threshold resonances on reaction rates in 
H$_2$+Cl and H$_2$+F \cite{NTR-PRA,Simbotin2015} for the ultracold case 
where only the $s$-wave contribution plays a role. 
In this article, we extend our analysis to higher partial waves; for $\ell \geq 2$, we
employ a model based on three coupled open channels that incorporates the 
key ingredients while allowing for easy tuning of the resonances for each partial wave
$\ell$. We uncover specific scaling differing from the expected Wigner's scaling laws 
for given partial waves $\ell$. 

We first review scattering theory for multi-channel problems to establish the notation, 
followed by the corresponding Jost function treatment. Using mass scaling, we demonstrate
how near threshold resonances (NTRs) manifest themselves in H$_2$+Cl ($\ell=0$)
and H$_2$+D ($\ell=1$). We build on the Jost function approach to formulate the 
NTR regime scaling laws, and illustrate them with the simpler model for $\ell \geq 2$.
We finally discuss briefly the effect of a power-law potential tail on those scalings.

\section{Scattering}

In a multichannel scattering problem, the scattering wave function $\psi^{+}_{\bm k}$ for 
in incident projectile with momentum ${\bm p} =\hbar {\bm k}$ can be expanded onto a 
complete basis representing the channels. Here, we review the case of non-reactive 
processes, where the initial arrangement remains intact after the scattering event,
though the general results are applicable to the reactive case. 
A more thorough discussion can be found in several scattering textbooks, such 
as \cite{Taylor,Newton,Joachain}.

Below, we limit our discussion to the non-reactive case (although the results can be 
generalized to the rearrangement case \cite{Taylor}), and assume the
Hamiltonian
\begin{equation}
   \hat{H} = \left(\frac{{\bm p}^2}{2\mu} +  \hat{H}_{\rm target}\right) + V({\bm r},{\bm s}) \;, 
   \label{eq:H(total)}
\end{equation}
where $\bm r$ is the position projectile and $\bm s$ the set of coordinates describing the target, 
$\mu$ is the reduced mass of the projectile and target, and $V({\bm r},{\bm s})$ the interaction
between them. Here, $\hat{H}_{\rm target}$ dictates the target dynamics, with
$\hat{H}_{\rm target} \phi_n(\bm s) = E_n \phi_n(\bm s)$,
We expand
$\psi^{+}_{\bm k}(k,{\bm r},{\bm s})$ in the basis of $\phi_n(\bm s)$, {\it i.e.}
\begin{equation}
   \psi^{+}_{\bm k}(k,{\bm r},{\bm s}) = \sum_{\;\;n}\hspace{-.20in}\int \;\;
   \eta_n({\bm r}) \phi_n(\bm s) \;,
   \label{eq:psi+expansion}
\end{equation}
where $\eta_n({\bm r})$ correspond to the channel wave functions, and the
sum runs over discrete and continuum states.
Their asymptotic form is written as
\begin{equation}
   \eta_n({\bm r}) \xrightarrow{{\bm r}\rightarrow\infty} 
   \delta_{ni} e^{i{\bm k}\cdot {\bm r}}
   + f_{ni} \frac{e^{i k_n r}}{r} \; ,
\label{eq:wf-multi-eta}
\end{equation}
where $ f_{n i}\equiv f({\bm k}_n, n \leftarrow {\bm k},i) $ stands for the
scattering amplitude from the initial/incident channel $i$ with momentum $ {\bm p}=\hbar {\bm k}$ into the channel $n$ with momentum $ {\bm p}_n=\hbar{\bm k}_n$.
 
Applying $\hat{H}$ onto the expansion (\ref{eq:psi+expansion}) for 
$ \psi^{+}_{\bm k}(k,{\bm r},{\bm s})$, and using orthonormality 
$\int d{\bm s} \phi^*_{m}(\bm s) \phi_{n}(\bm s) = \delta_{mn}$,
of the basis $\phi_{n}$, one gets \cite{Taylor}

\begin{equation}
   -\frac{\hbar^2\nabla^2}{2\mu} \eta_{m}({\bm r}) 
   + \sum_{n}\hspace{-.21in}\int \; V_{mn}({\bm r})\eta_{n}({\bm r})
= (E-E_{m} ) \eta_{m} ({\bm r}) \;,
\label{eq:multi-channel-full-coupled}
\end{equation}
where
\begin{equation}
   V_{mn}({\bm r}) \equiv \int d{\bm s}\; \phi^*_{m}(\bm s)  V({\bm r},{\bm s})  \phi_{n}(\bm s) \;.
\end{equation}
Although there is in principle an infinite set of coupled equations due to the infinite number 
of states arising from the continuum, one generally neglects its contribution and restricts
the number $N$ of discrete terms considered to obtain {\it the close-coupling approximation}.





In that case, if we label the initial/incident channel by $i=1$, the solution
${\bm \eta}_1(\bm r)$ of the scattering problem, Eq.(\ref{eq:multi-channel-full-coupled}),
can be rewritten in the matrix form
\begin{equation}
   \nabla^2 {\bm \eta}_1(\bm r) - \mtrx{U}(\bm r)  {\bm \eta}_1(\bm r) + \mtrx{K}^2  {\bm \eta}_1(\bm r)  =0 \;,
\label{eq:multi-channel-eta_1}
\end{equation}
where 
\begin{equation}
   {\bm \eta}_1(\bm r) = \left( \begin{array}{c}  \eta_{i=1} (\bm r )  \\ \vdots \\  \eta_N (\bm r) \end{array} \right) \;\;, \;\;
   \mtrx{K} = \left( \begin{array}{ccc} k_1 & & 0 \\ & \ddots & \\ 0 & & k_N \end{array} \right)
   \;,
   \label{eq:eta-matrices}
   \end{equation}
with $k_n = \sqrt{2\mu (E-E_n )/\hbar^2}$,
and where $\mtrx{U}(\bm r)$ is a $N\times N$ matrix with elements 
$U_{mn} = \frac{2m}{\hbar^2} V_{mn}$. 
In general, the incident wave
can be in any of the channels $n= 1, \dots , N$, leading to $N$ distinct solutions 
${\bm \eta}_1(\bm r), \dots ,{\bm \eta}_N(\bm r)$, where each 
${\bm \eta}_n(\bm r)$ describes a collision beginning
in channel $i=n$, so that Eq.(\ref{eq:multi-channel-eta_1}) can be rewritten as
\begin{equation}
   \nabla^2  \overline{\bm \eta}(\bm r) - \mtrx{U}(\bm r)   \overline{\bm \eta}(\bm r) 
   + \mtrx{K}^2  \overline{\bm \eta}(\bm r)  =0 \;,
\label{eq:multi-channel-eta}
\end{equation}
where $\overline{\bm \eta}(\bm r)$ is a matrix with each column being the solution for an initial channel
\begin{equation}
 \overline{\bm \eta}(\bm r) = ( {\bm \eta}_1(\bm r) , {\bm \eta}_2(\bm r), \dots , {\bm \eta}_N(\bm r) ) \;,
 \label{eq:eta-matrix}
\end{equation}
with
\begin{equation}
  \!\!\!{\bm \eta}_1 \!=\! \left( \begin{array}{c}  \eta_{i=1}  \\ \eta_{2}  \\ \vdots \\  \eta_N \end{array} \right) \!,  
  {\bm \eta}_2\!=\!  \left( \begin{array}{c}  \eta_{1}  \\ \eta_{i=2} \\ \vdots \\  \eta_N \end{array} \right)\!,  \dots ,
  {\bm \eta}_N \!=\!  \left( \begin{array}{c}  \eta_1 \\ \eta_{2} \\ \vdots \\  \eta_{i=N} \end{array} \right)\!.
\end{equation}

We consider the case where the system is rotationally
invariant and spinless, so that the solutions ${\bm \eta}_n(\bm r)$  can be written in
a partial wave expansion of the form \cite{Taylor}
\begin{equation}
   {\bm \eta}_n (\bm r) = \sum_{\ell =0}^{\infty} \frac{(2\ell +1 )}{kr} {\bm \psi}^{(\ell)}_n (r) 
   P_\ell (\cos\theta) \;,
   \label{eq:eta-rotational-invariant-spinless}
\end{equation}
which satisfies the matrix radial equation
\begin{equation}
   \left[ \mtrx{I} \frac{d^2}{dr^2} - \mtrx{I} \frac{\ell(\ell +1 )}{r^2} -  \mtrx{U}(r)  + \mtrx{K}^2 \right]  
   {\bm \psi}^{(\ell)}_n (r) = 0 \;.
   \label{eq:SE-rotational-invariant-spinless}
\end{equation}
The vector ${\bm \psi}^{(\ell)}_n (r)$ is the radial solution with the incident wave in 
channel $i=n$, and for each angular momentum $\ell$, there are $N$ distinct radial 
functions ${\bm \psi}^{(\ell)}_n (r)$. Each vector ${\bm \psi}^{(\ell)}_n (r)$ has $N$ 
components $\psi^{(\ell)}_{m  n}(r)$, and their asymptotic form is given by 
\begin{eqnarray}
   \psi^{(\ell)}_{m  n}(r) & \xrightarrow{r\rightarrow\infty} &
   C(k_n) \left[  i^\ell s_\ell (k_n r ) \delta_{m  n} +  k_n f^{(\ell )}_{m  n} e^{ik_{m}r}\right] 
   \nonumber \\
  && = C(k_n) \frac{i^{2\ell +1}}{2} \Bigg[ e^{-ik_{n }r}  \delta_{mn} 
  \nonumber \\ && \hspace{.35in}
      - (-1)^\ell \sqrt{\frac{k_n}{k_{ m}}} S^{(\ell)}_{mn} e^{ik_{ m}r} \Bigg] \;,
      \label{eq:single-channel-f-S}
\end{eqnarray}
where $s_\ell (x)= x j_\ell (x)$ is the Riccati-Bessel function, $C(k_n)$ is a normalization constant, and where $f^{(\ell )}_{mn}$ and 
$S^{(\ell )}_{mn}$ are related by
\begin{equation} 
   S^{(\ell )}_{mn} = \delta_{mn}+2i\sqrt{k_{m} k_n} f^{(\ell )}_{mn} \;. 
   \label{eq:multi-S_l-f_l} 
\end{equation}
The $\sqrt{k_n / k_{m}}$ factor appearing with $S^{(\ell)}_{mn}$ in
Eq.(\ref{eq:single-channel-f-S}) ensures the unitarity of the S-matrix.

Regrouping all vectors ${\bm \psi}^{(\ell)}_n (r)$ into a single $N\times N$ matrix 
as for $\overline{\bm \eta}(\bm r)$ in Eq.(\ref{eq:eta-matrix}), we have
\begin{equation}
 \overline{\bm \Psi}^{(\ell )}(r) = ( {\bm \psi}^{(\ell)}_1 (r) , {\bm \psi}^{(\ell)}_2 (r), \dots , 
               {\bm \psi}^{(\ell)}_N (r) ) \;,
 \label{eq:Psi-matrix}
\end{equation}
and the asymptotic forms in Eq.(\ref{eq:single-channel-f-S}) are rewritten as
\begin{eqnarray}
 \overline{\bm \Psi}^{(\ell )}(r)  &\!  \xrightarrow{r\rightarrow \infty}  &
 \left[ i^\ell s_\ell (\mtrx{K} r) +  e^{i\mtrx{K}r}\mtrx{F}^{(\ell)} \mtrx{K} \right]  C(\mtrx{K}) \nonumber \\
 & & \hspace{-.9in} = \frac{i^{2\ell +1}}{2}\! \left[ e^{-i\mtrx{K}r} \!-\! (-1)^\ell e^{i\mtrx{K}r} 
        \mtrx{K}^{-1/2}\mtrx{S}^{(\ell)} \mtrx{K}^{1/2} \right] \! C(\mtrx{K}) ,
   \label{eq:asymp-matrix}
\end{eqnarray}
where  $\mtrx{F}^{(\ell)} $ and $\mtrx{S}^{(\ell)} $ are the matrices for 
$f^{(\ell )}_{mn}$ and $S^{(\ell )}_{mn}$,  $\mtrx{K}$ given in Eq.(\ref{eq:eta-matrices}) 
with $\mtrx{K}^{\pm1/2} = {\rm diag}\{k_j^{\pm1/2}\}$, and with the various diagonal 
matrices defined as $C(\mtrx{K})= {\rm diag}\{C(k_j)\}$, 
$s_\ell (\mtrx{K} r) = {\rm diag}\{s_\ell (k_j r)\}$, and 
$e^{\pm i\mtrx{K}r} ={\rm diag}\{ e^{\pm ik_j r} \}$.
Eq.(\ref{eq:multi-S_l-f_l}) is then written as
\begin{equation}
   \mtrx{S}^{(\ell)}= \mtrx{I} + 2i \mtrx{K}^{1/2}\mtrx{F}^{(\ell)} \mtrx{K}^{1/2}\;.
\end{equation}
The order of the matrix multiplication is important in those matrix forms.

Differential cross sections for multi-channel scattering with and without rearrangement  
are given by \cite{Taylor}
\begin{equation}
   \frac{d\sigma_{m\leftarrow n}}{d\Omega} = \frac{k_{m}}{k_n} | f_{mn}|^2\; ,
\end{equation}
In general, the exact form 
of the expression depends on the angular momenta, internal structure, and the 
exact interactions entering the Hamiltonian (such as interaction with external fields, etc.).
For the simpler rotationally invariant and spinless system satisfying Eqs.(\ref{eq:eta-rotational-invariant-spinless}) and (\ref{eq:SE-rotational-invariant-spinless}), we have
\begin{equation}
   f_{m n} = \sum_{\ell =0}^\infty (2\ell + 1) f^{(\ell )}_{m n} P_\ell (\cos\theta)\;.
\end{equation}
Using the properties of Legendre polynomials $P_\ell$, and after integration over
angles, one gets
\begin{equation}
   \sigma_{m \leftarrow n}(k_n ) = 4\pi   \frac{k_{m}}{k_n}   
   \sum_{\ell =0}^\infty (2\ell + 1) |f^{(\ell )}_{mn} |^2 \;,
   \label{eq:multi-sigma-f}
\end{equation}
which can be rewritten, with the help of Eq.(\ref{eq:multi-S_l-f_l}), as
\begin{eqnarray}
   \sigma_{m \leftarrow n}(k_n ) & = & \frac{\pi}{k_n^2}  
   \sum_{\ell =0}^\infty (2\ell + 1) |\delta_{mn} - S^{(\ell )}_{mn} |^2 \;, 
   \label{eq:multi-sigma-S} \\
   & = &  \frac{\pi}{k_n^2} 
   \sum_{\ell =0}^\infty (2\ell + 1) |T^{(\ell )}_{mn} |^2 \;,\\
   & = & \sum_{\ell =0}^\infty (2\ell + 1)  \sigma^{(\ell)}_{m \leftarrow n}(k_n ) \;,
   \label{eq:multi-sigma-T}
\end{eqnarray}
where we define the partial cross section $ \sigma^{(\ell)}_{m \leftarrow n}(k_n )$
in term of the T-matrix, namely
\begin{equation}
    \sigma^{(\ell)}_{m \leftarrow n}(k_n ) \equiv \frac{\pi}{k_n^2}|T^{(\ell )}_{mn} |^2 \;.
   \label{eq:sigma_ell-T}
\end{equation}
The T-matrix $T^{(\ell )}_{mn} = \delta_{mn} - S^{(\ell )}_{mn}$ can be 
written as
\begin{equation}
   \mtrx{T} = \mtrx{I} - \mtrx{S} \;.
\end{equation}
Using the unitarity of the S-matrix, namely 
$1 = \sum_{m} |S^{(\ell )}_{mn}|^2 = |S^{(\ell )}_{nn}|^2 + \sum_{m \neq n} |S^{(\ell )}_{mn}|^2$, 
the elastic and total inelastic cross sections are simply
\begin{eqnarray}
   \sigma_{n}^{\rm elas} & \equiv \sigma_{n \leftarrow n} & \!\!\!= \frac{\pi}{k_n^2}
   \sum_{\ell =0}^\infty (2\ell + 1) |S^{(\ell )}_{n n}|^2 \;, \\
   \sigma_{n}^{\rm inel} &\!\! \equiv \displaystyle \sum_{m\neq n} \sigma_{m \leftarrow n} 
   &\!\! \!= \frac{\pi}{k_n^2} \sum_{\ell =0}^\infty (2\ell + 1) [1-|S^{(\ell )}_{n n}|^2] .
\end{eqnarray}
We note that in the zero-energy limit, the cross sections are given by the $s$-wave ($\ell=0$)
partial wave and can be expressed in terms of a complex scattering length 
$a_n =\alpha_n -i\beta_n$ \cite{Balakrishnan1997b,NTR-PRA}, namely 
$\sigma_{n}^{\rm elas}\sim 4\pi|a_n|^2$ and $\sigma_{n}^{\rm inel} \sim 4\pi \beta_n/k_n$,
which exemplify the usual Wigner's threshold regime 
\cite{Balakrishnan1997b,NTR-PRA,Wigner1948} 
.

Resonances can be understood from the appearance of poles in the structure of the S-matrix, 
and the proximity of the scattering energy $E$ (or momentum $k$) from these poles.
A useful formalism to explore these effect is based on the Jost function.

\section{Jost function}
\label{sec:Jost-multi}

We consider rotationally invariant and spinless systems.
For $N$ coupled channels, the regular solution  $\overline{\bm \phi}^{(\ell)}(r)$ is an
$N\times N$ matrix with elements $\phi^{(\ell )}_{mn}$ satisfying the system of
coupled radial equations,
\begin{equation}
   \mtrx{I} \frac{d^2}{dr^2} \overline{\bm \phi}^{(\ell)}(r) = \left[  \mtrx{U}(r) + \mtrx{I} \frac{\ell(\ell +1 )}{r^2} - 
   \mtrx{K}^2 \right]  \overline{\bm \phi}^{(\ell)}(r)  \;.
\end{equation}
The element $\phi^{(\ell )}_{mn}$ must satisfy the boundary condition 
$\phi^{(\ell )}_{mn}(r) \sim \delta_{mn}(k_n r)^{\ell +1}$ as $r\rightarrow 0$.
For potential elements $U_{mn}$ less singular than $r^{-2}$ at the origin and vanishing 
faster than $r^{-3}$ at $\infty$, the asymptotic behavior 
of $\phi^{(\ell )}_{mn}$ can be written as
\begin{eqnarray}
  \phi^{(\ell )}_{mn}(r) & \xrightarrow{r\rightarrow 0}
   & \delta_{mn} s_{\ell} (k_{m} r)  \;,  \label{eq:many-ell-asy-r=0} \\
  \phi^{(\ell )}_{mn}(r) & \xrightarrow{r\rightarrow \infty}
   & s_{\ell} (k_{m} r) A^{(\ell)}_{mn}
   + c_{\ell} (k_{m}r) B^{(\ell)}_{mn} \;.
     \label{eq:many-ell-asyAB}
\end{eqnarray}
The asymptotic $\phi^{(\ell )}_{mn}(r)$ at large
$r$ can be written in terms of the free solutions $e^{\pm ikr}$, namely
\begin{eqnarray}
   \phi^{(\ell )}_{mn}(r) & \!\!\xrightarrow{r\rightarrow \infty}& \frac{i}{2}
   \left[ (A^{(\ell)}_{mn} -iB^{(\ell)}_{mn} ) e^{- i(k_{m} r-\ell\pi/2)} \right. \nonumber \\ &&\left.
   \hspace{.05in} - (A^{(\ell)}_{mn} +iB^{(\ell)}_{mn} )e^{+i(k_{ m} r- \ell\pi/2)} \right] ,\\ 
   & \hspace{-.35in} \equiv & \hspace{-.35in}\frac{i^{\ell +1}}{2} \!
   \left[ \mathcal J^{(\ell)}_{mn}  e^{- ik_{m } r} 
   \!-\! (-1)^\ell \mathcal J^{(\ell)*}_{mn} e^{+ik_{m } r}  \right] ,
   \label{eq:multi-asyF}
\end{eqnarray}
where, we define the Jost matrix element as
\begin{equation}
\label{eq:multi-J-def}
  \mathcal J^{(\ell)}_{mn}  \equiv A^{(\ell)}_{mn}
  - iB^{(\ell)}_{mn} \; .
\end{equation}
or, in matrix form,
\begin{equation}
  \overline{\bm \phi}^{(\ell)}(r)   \xrightarrow{r\rightarrow \infty} \frac{i^{\ell +1}}{2}
   \left[ e^{-i\mtrx{K}r} \bm{\mathcal{J}}_{\ell} 
   - (-1)^{\ell} e^{i\mtrx{K}r} \bm{\mathcal{J}}_{\ell}^* \right]\;.
\end{equation}
Multiplying $\overline{\bm \phi}^{(\ell)}$ by 
$ i^\ell \bm{\mathcal{J}}_{\ell}^{-1} C(\mtrx{K})$ 
on the right, we find by comparing with the physical solution given by Eq.(\ref{eq:asymp-matrix}),
\begin{equation}
    \overline{\bm \Psi}^{(\ell )}= i^{\ell}\overline{\bm \phi}^{(\ell)}\bm{\mathcal{J}}_{\ell}^{-1}C(\mtrx{K}) \;,
\end{equation}
with
\begin{equation}
   \mtrx{S}^{(\ell)} = \mtrx{K}^{1/2} \bm{\mathcal{J}}_\ell^*  \bm{\mathcal{J}}_\ell^{-1}  \mtrx{K}^{-1/2} 
\end{equation}

If $\ell$ is not a good quantum number, for example due to interactions 
with external fields, the Jost-matrix can still be written in term of two matrices \cite{Taylor}
as in Eq.(\ref{eq:multi-J-def})
\begin{equation}
  \bm{\mathcal{J}} \equiv \mtrx{A} - i \mtrx{B} \;.
\end{equation}
and the matrix expression relating the S-matrix and the Jost-matrix is still valid 
\begin{equation}
   \mtrx{S} = \mtrx{K}^{1/2} \bm{\mathcal{J}}^*  \bm{\mathcal{J}}^{-1}  \mtrx{K}^{-1/2} \;.
   \label{eq:S-matrix-def}
\end{equation}
The inverse of the Jost-matrix in $\mtrx{S}$ is given by
\begin{equation}
    \bm{\mathcal{J}}^{-1} = \frac{1}{\det(\bm{\mathcal{J}})} 
   \left[ \mathrm{Cof}(\bm{\mathcal{J}}) \right]^T,
   \label{eq:J-inv}
\end{equation}
with $\left[\mathrm{Cof}(\bm{\mathcal{J}}) \right]^T$ the transpose of the matrix of cofactors
of $\bm{\mathcal{J}}$ and $\det(\bm{\mathcal{J}})$ the determinant of $\bm{\mathcal{J}}$.
These expressions are particularly instructive to understand the effect a Near Threshold 
Resonance (NTR) on the scattering cross sections. 

\section{Near Threshold Resonances}

We consider resonances occurring due to existence of a quasi-bound state in the entrance channel
of a scattering system. We first look at few examples in benchmark chemical reactions, explain
the energy scaling due to these near threshold resonances based on the properties of the Jost functions, and employ a three-open channel model to illustrate the effect for higher partial
waves $\ell$ up to $f$-wave ($\ell=3$). We end the discussion with the elastic case, where
additional considerations on the range of the potential affect the validity of our results.

\subsection{Benchmark reactions}

\begin{figure}[ht]
\includegraphics[clip,width=.42\linewidth]{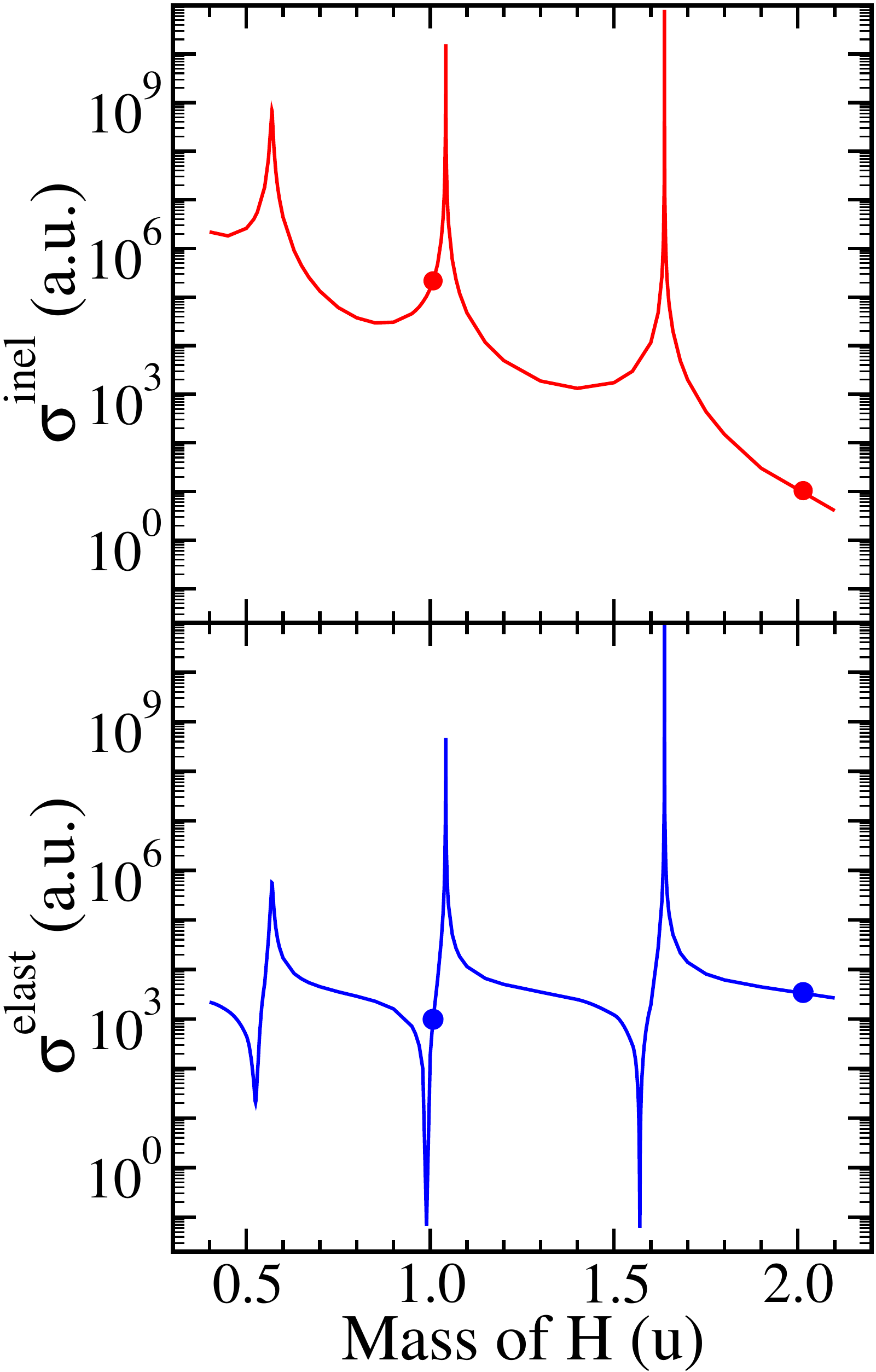}
\includegraphics[clip,width=.48\linewidth]{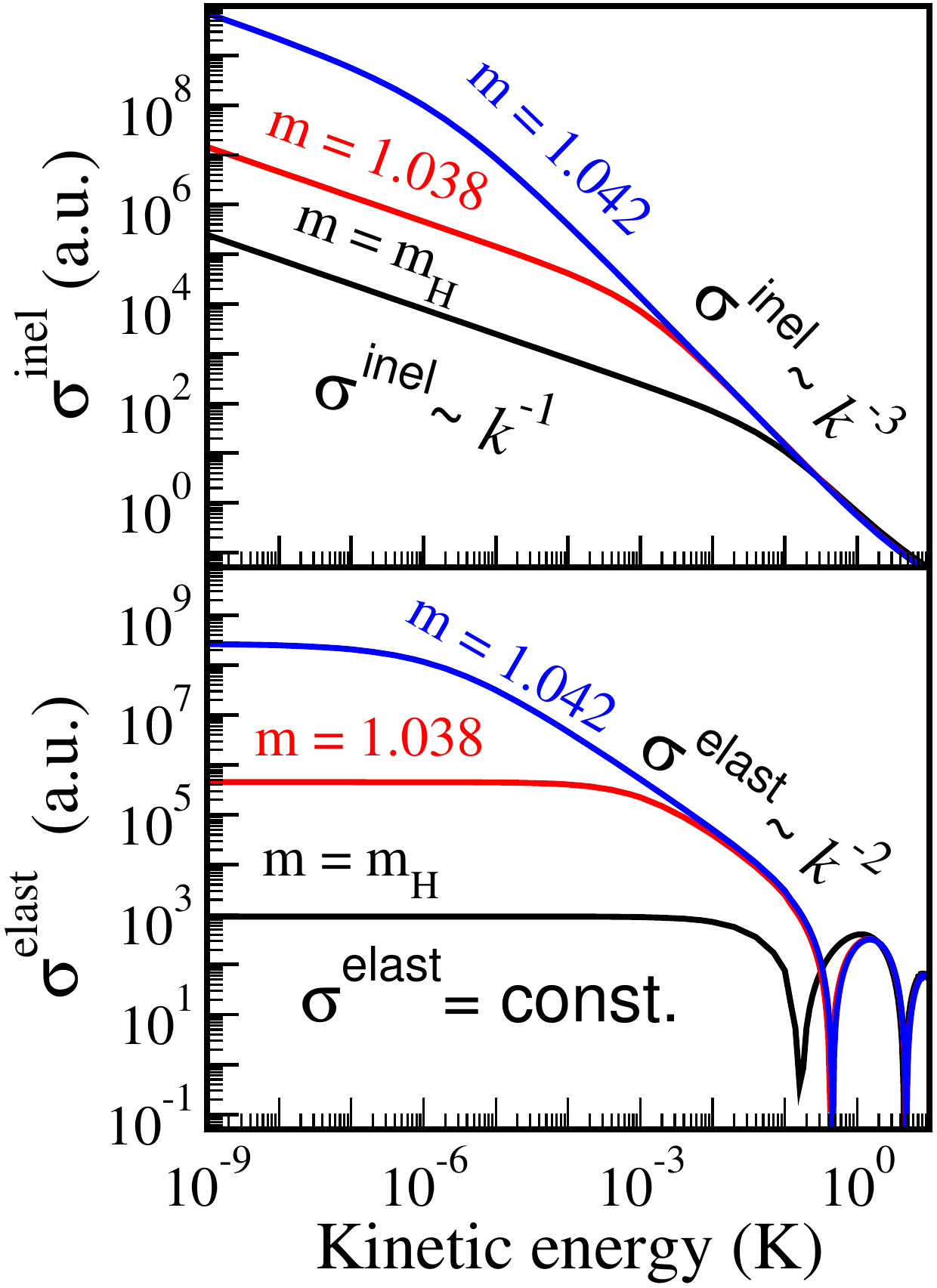}
\caption{Left panel: elastic and total inelastic cross sections vs. $m$ (the mass of H, with
  circles stand for the true mass of H and D) for 
  the entrance channel H$_2(v\!\!=\!\!1,\,j\!\!=\!0)$ + Cl at a collision energy 
  $E/k_B = 1$ nK. Right panel: corresponding energy dependence of the inelastic 
  (a) and elastic (b) cross sections for $m=1.0078\,\mathrm{u}=m_{\rm{H}}$ 
  (true mass), 1.038\,u, and 1.042\,u (blue). 
}
\label{fig:alpha-beta+CX}
\end{figure}

In our previous work \cite{NTR-PRA,Simbotin2015}, we have used two
benchmark chemical reactions to investigate near threshold resonances in the ultracold
temperature regime. In particular, we studied H$_2$+Cl and H$_2$+F reactions in the
$s$-wave ($\ell =0$) scattering regime, by varying the mass of H as suggested
originally by \cite{Bodo:JPB:2004}, which changes the
relative position of channels and bound and quasi-bound states, allowing to theoretically 
tune resonances and scattering processes. This approach is
similar to modifying the PES itself~\cite{JMH:JCP:2007}.
The results presented here were obtained using the \textsc{abc} reactive scattering code of 
Manolopoulos and coworkers~\cite{ABC:CPC:2000}, which has been optimized 
for ultralow energies in previous studies of H$_2$+D \cite{PCCP-H2+D},
and H$_2$+Cl \cite{NTR-PRA,Simbotin2015},

The first case we discuss is H$_2$+Cl for $s$-wave scattering, for which
the details of the calculations and mass scaling
are given in \cite{NTR-PRA,Simbotin2015}.
This system was recently investigated at ultralow temperatures 
\cite{Bala2012,NTR-PRA,Simbotin2015}; we used 
the potential energy surface (PES) developed by Bian and Werner~\cite{Bian-H2Cl}.
Fig.~\ref{fig:alpha-beta+CX} shows the
results for H$_2$+Cl, with the left panels depicting the elastic $\sigma_{n}^{\rm elas}$    
and total inelastic $\sigma_{n}^{\rm inel}$ cross sections for the initial channel 
$n$ corresponding to H$_2(v=1,j=0)$+Cl as a function of the mass of H, $m$, at
a collision energy 1 nK. It shows sharp variations of the cross sections for
specific values of $m$: the real mass of H ($m_{\rm H}$) and D ($m_{\rm D}$)
are indicated by solid circles.
The right panels illustrate the effect of near threshold resonances (NTRs) on 
cross sections for three values of $m$
starting from $m_{\rm H}$ and getting closer to the resonance shown on the left panels.
The cross sections are given
as a function of the scattering energy $E=\hbar^2k^2/2\mu$ defined from the threshold
of the entrance channel. 

As $k \rightarrow 0$, $\sigma_{n}^{\rm inel}$  reaches
the Wigner's regime, scaling as $k^{-1}$ for all three masses. 
For masses closer to the resonance, the scaling changes to $k^{-3}$
at higher energies (still ultracold and described by $\ell=0$ scattering only). 
This behavior appears to be universal; at higher energies (but still ultracold), 
$\sigma_{n}^{\rm inel}$ has the same value until it deviates from the universal 
NTR $k^{-3}$ scaling to join the Wigner $k^{-1}$ scaling at lower k. 
The elastic cross sections $\sigma_{n}^{\rm elas}$ is also shown for the same masses ;
the Wigner regimeÕs constant cross section as $k\rightarrow 0$ changes
to the expected $k^{-2}$ scaling for m near a resonance.

\begin{figure}[ht]
\centerline{\includegraphics[clip,width=1.0\linewidth]{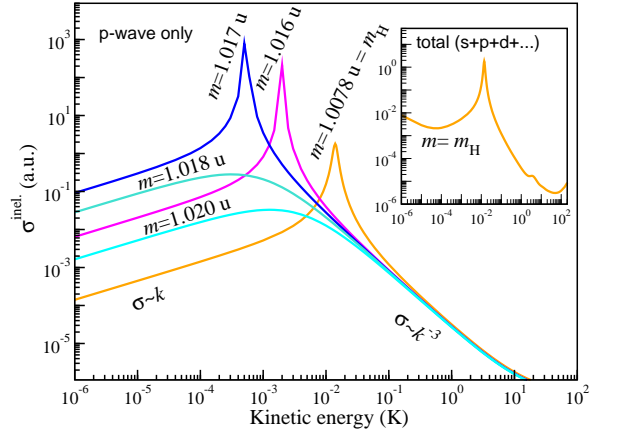}}
\caption{The $p$-wave contribution for the total inelastic (including reaction)
cross section for D + H$_2(v=1,j=0)$ for different values of $m$, as
indicated.  The inset shows the fully converged cross section,
including partial waves $\ell=0,1,2,3,4,\ldots$.
}
\label{fig:p-NTR}
\end{figure}

Similar behaviors were found for H$_2$+F in the $s$-wave ($\ell=0$) 
regime \cite{Simbotin2015} using the PES developed by 
Stark and Werner~\cite{SWpes}: this system has also been studied in 
the ultracold regime by \cite{H2F-Bala,Bodo:JPB:2004}.
The origin of the NTR regime for $\ell=0$ was 
investigated in our previous work, first as the result of the proximity of a pole of 
the S-matrix to the real positive $k$-axis \cite{NTR-PRA}, and then in terms of the
Jost function \cite{Simbotin2015}. More recently, we have explored other
benchmark reactions involving hydrogen and its isotopes, namely H$_2$+D 
\cite{NJP-para-ortho-H2+D} and D$_2$+H \cite{D2+H-arXiv} using the PES 
of \cite{bkmp2:jcp96}, to probe the effect of the nuclear
spin symmetry on the scattering processes. In particular, we identified a $p$-wave
($\ell =1$) shape resonance in the H$_2(v=1,j=0)$+D entrance channel for para-H$_2$.
The details of the calculations are given in \cite{NJP-para-ortho-H2+D}. In Fig.~\ref{fig:p-NTR} (inset),
we show the total inelastic cross section vs $E$ for the real mass of H, with the leading 
contribution of $s$-wave ($\ell=0$) scattering as $E\rightarrow 0$, the dominating
contribution of the $p$-wave ($\ell=1$) resonance near $E/k_B\sim 100$ mK, the
smaller structure due to the $d$-wave ($\ell=2$) near 5 K, and higher partial wave 
contributions at larger energies still.  

The main plot of Fig.~\ref{fig:p-NTR} depicts
the $p$-wave contribution only, for different value of $m$ (mass of H).  As $m$
increases from the real mass of H ($m_{\rm H}$), the position of the resonance 
shifts to lower energies and is accompanied by an increased magnitude until it 
disappears near $m\sim 1.018$ u, at which point the van der Waals complex
H$_2\cdots$D acquires a new bound state. As $m$ increases further, the 
maximum in the cross section starts shifting to larger energies with a 
decreasing magnitude. We note that using Rydberg-dressed interactions between
H$_2$ and D mimics the variation of the mass of H by changing the amount
of Rydberg admixing in the colliding partners \cite{Rydberg-tuning}. Fig.~\ref{fig:p-NTR} 
clearly points to two $k$-scaling regimes on either side of the resonance, the
expected $k$ Wigner scaling as $k\rightarrow 0$, and a different $k^{-3}$ NTR scaling
at higher $k$. We also note that while the Wigner regime tends to different values (all
with the $k$-dependence), the total inelastic cross section ($p$-wave) coalesces on
a single curve in NTR regime, in a fashion similar to the $s$-wave NTR regime discussed
above and shown in Fig.~\ref{fig:alpha-beta+CX}.

In the next section, we give a theoretical framework based on Jost functions, and generalize
the treatment to any partial wave $\ell$.

\subsection{Jost function theoretical framework: inelastic processes}
 
To explain the appearance of the NTR regime, we turn to the properties
of the $S$-matrix. From Eq.(\ref{eq:S-matrix-def}),   
$\bm{\mathsf{S}} = \bm{\mathsf{K}}^{1/2} \bm{\mathcal{J}}^*  \bm{\mathcal{J}}^{-1} 
\bm{\mathsf{K}}^{-1/2}$,and Eq.(\ref{eq:J-inv}), 
$\bm{\mathcal{J}}^{-1} = \left[ \mathrm{Cof}(\bm{\mathcal{J}}) \right]^T/\det(\bm{\mathcal{J}})$, 
we can rewrite the matrix element $S^{(\ell)}_{fi}$ for the partial cross section 
$\sigma^{(\ell )}_{f\leftarrow i}$ in Eq.(\ref{eq:sigma_ell-T}) as
\begin{equation*}
   S_{fi} = \sqrt{\frac{k_f}{k_i}} \left(\bm{\mathcal{J}}^* 
   \frac{\left[ \mathrm{Cof}(\bm{\mathcal{J}}) \right]^T}
   {\det(\bm{\mathcal{J}})}\right)_{fi} = \sqrt{\frac{k_f}{k_i}} 
   \frac{\sum_j {\cal J}_{fj}^* C_{ij}}{\det(\bm{\mathcal{J}})} \;,  
\end{equation*}
where we omit $(\ell)$ for simplcity, and $C_{ji}=\mathrm{Cof}(\bm{\mathcal{J}})_{ji}$ 
is the cofactor of  $\bm{\mathcal{J}}$.
Adding and subtracting ${\cal J}_{fj}$ in the sum, and with 
${\cal J}_{fj}^*-{\cal J}_{fj}=2iB_{fj}$, we
have ${\cal J}_{fj}^* C_{ij}= {\cal J}_{fj} C_{ij} + 2i B_{fj} C_{ij}$, so that
\begin{eqnarray} 
   \sum_j {\cal J}_{fj}^* C_{ij} &= & \sum_j {\cal J}_{fj} C_{ij} + 2i \sum_j B_{fj} C_{ij} \nonumber \\
        & = & \delta_{fi}\det(\bm{\mathcal{J}}) +  2i \sum_j B_{fj} C_{ij} \;,
\end{eqnarray}   
where we use the properties of a determinant in term of cofactors.
The matrix element $S_{fi}$ is then
\[ S_{fi} =   \delta_{fi} + \sqrt{\frac{k_f}{k_i}}\frac{2i \sum_j B_{fj} C_{ij}}{\det(\bm{\mathcal{J}})} \;,  \]
and the element $T_{fi}$ of the T-matrix $\bm{\mathsf{T}}=\bm{\mathsf{1}}-\bm{\mathsf{S}}$ 
simply becomes 
\[ T_{fi} =  -2i \sqrt{\frac{k_f}{k_i}} \frac{\sum_j B_{fj} C_{ij}}{\det(\bm{\mathcal{J}})} \;. \]
The partial cross section $\sigma^{(\ell )}_{f\leftarrow i}$ defined in Eq.(\ref{eq:sigma_ell-T}) is \begin{equation}
   \sigma_{f\leftarrow i} = \frac{\pi}{k_i^2} |T_{fi}|^2 
   = \frac{4\pi k_f}{k_i^3} \frac{|\sum_j B_{fj} C_{ij}|^2}
   {|\det(\bm{\mathcal{J}})|^2}\;.
\end{equation}
The resonance being due to a quasi-bound state in the entrance channel $i$, we isolate its
contribution and write the determinant as
\begin{equation} 
   \det(\bm{\mathcal{J}}) = \sum_n {\cal J}_{in} C_{in}= C_{ii}({\cal J}_{ii} + {\sf j}_{ii} ) \;, 
\end{equation}
where
\begin{equation}
    {\sf j}_{ii} \equiv \frac{1}{C_{ii}}\sum_{n\neq i}  {\cal J}_{in} C_{in} \;. 
    \label{eq:def-det-J}
\end{equation}
The denominator $|\det(\bm{\mathcal{J}})|^2$ of $\sigma_{f\leftarrow i}$ becomes
\begin{equation}  
   |\det(\bm{\mathcal{J}})|^2 = | C_{ii}|^2 |D|^2 \;,  \mbox{ where } D \equiv {\cal J}_{ii} + {\sf j}_{ii}  \;.
   \label{eq:def-det-D}
\end{equation}   
As the resonance is in the entrance channel $i$, the cofactor $C_{ii}$ includes all but 
the entrance channel and is a well-behaved function almost independent of $k_i$ as 
$k_i\rightarrow 0$ \cite{Simbotin2015}. The effect of the resonance is accounted for 
in $D$ (mostly via the ${\cal J}_{ii}$ contribution since ${\sf j}_{ii}$ includes all other channels).
For clarity, we label the entrance channel defining the threshold for the scattering energy by
$i=1$, and simply adopt the notation $k_i=k_1$ and $\ell_i=\ell_1$, so that
\begin{equation} 
   \sigma_{f\leftarrow 1} = \frac{4\pi k_f}{k_1^3} \frac{|\sum_j B_{fj} C_{1j}|^2}
   { | C_{11}|^2 |D|^2} \;. 
   \label{eq:sigma_inel_CD}
\end{equation}   
To understand the behavior of $\sigma_{f\leftarrow 1}$ at small $k_1$, we examine the 
$k_1$-dependence of $B_{mn}(k_1)$ and $C_{mn}(k_1)$ starting from that of 
$A_{mn}(k_1)$ and $B_{mn}(k_1)$
extracted from Eqs.(\ref{eq:many-ell-asy-r=0}) and (\ref{eq:many-ell-asyAB}). If none of the 
index include the initial channel, we find
\begin{equation}
  \left. \begin{array}{l} m\neq 1 \\ n\neq 1 \end{array} \right\} \Rightarrow 
  \left\{ \begin{array}{l} 
   A_{mn} = A_{mn}^{(0)} + A_{mn}^{(2)} k_1^2 + \dots , \\
   B_{mn} = B_{mn}^{(0)} + B_{mn}^{(2)} k_1^2 + \dots , 
   \end{array} \right.
   \label{eq:A_mn+B_mn}
\end{equation}
where $A_{mn}^{(s)}$ and $B_{mn}^{(s)}$ for various $s$ are constants. If both
$m$ and $n$ are equal to one, we have
\begin{equation}
  \left. \begin{array}{l} m=1 \\ n= 1 \end{array} \right\} \!\Rightarrow \!
  \left\{ \begin{array}{l} 
   A_{11} \!=\!  A_{11}^{(0)} + A_{11}^{(2)} k_1^2 + \dots , \\
   B_{11} \!=\! k_1^{2\ell_1 + 1}\! \left( \!B_{11}^{(0)} \!+\! B_{11}^{(2)} k_1^2 \!+\! \dots \right) , 
   \end{array} \right.
   \label{eq:A_11+B_11}
\end{equation}
while if only one index is equal to one, we find
\begin{equation}
  \left. \begin{array}{l} m\neq 1 \\ n= 1 \end{array} \right\} \!\Rightarrow \!
  \left\{ \!\begin{array}{l} 
   A_{m1}  \!=\!  k_1^{\ell_1 + 1} \! \left( \! A_{m1}^{(0)} \! + \! A_{m1}^{(2)} k_1^2 \!+\! \dots  \right) , \\
   B_{m1} \!=\!  k_1^{\ell_1 + 1} \! \left( \! B_{m1}^{(0)} \! + \! B_{m1}^{(2)} k_1^2 \!+\! \dots \right) , 
   \end{array} \right.
   \label{eq:A_m1+B_m1}
\end{equation}
and
\begin{equation}
  \left. \begin{array}{l} m=1 \\ n\neq  1 \end{array} \right\} \!\Rightarrow \! 
  \left\{ \! \begin{array}{l} 
   A_{1n} \!=\!   k_1^{-\ell_1 - 1}\!  \left( \! A_{1n}^{(0)} \! + \! A_{1n}^{(2)} k_1^2 \! +\!  \dots \!\right), \\
   B_{1n} \!=\!   k_1^{\ell_1} \! \left( \! B_{1n}^{(0)} \!+ \! B_{1n}^{(2)} k_1^2 \! +\!  \dots \right).
   \end{array} \right. \label{eq:A_1n+B_1n}
\end{equation}
From these, we get (with $f\neq 1$)
\begin{equation}
   \left. \begin{array}{ll} B_{11} & \sim k_1^{2\ell_1 + 1} \\ B_{f1} & \sim k_1^{\ell_1 + 1} 
   \\ B_{1j\neq 1} & \sim k_1^{\ell_1}  \\ B_{fj\neq 1} & \sim \mbox{const.}  \end{array} \!\right\}   
     ,\mbox{and} 
   \left. \begin{array}{l} C_{11}  \sim C_{11}^{(0)}= \mbox{const.}  \\ C_{1j\neq 1}  \sim C_{1j\neq 1}^{(0)} k_1^{\ell_1 + 1}\end{array}\! \right\}
   , \label{eq:C_ij}
\end{equation}
where both $C_{11}^{(0)}$ and $C_{1j\neq 1}^{(0)}$ are complex constant.
For inelastic processes ($f\neq 1$), these give $\sum_j B_{fj} C_{1j} \sim k_1^{\ell_1 + 1}$, 
so that, together with $k_f$ reaching a finite value as $k_1\rightarrow 0$, we obtain
\begin{equation} 
   \sigma^{\rm inel}_{f\leftarrow 1} \sim  \frac{k_f}{k_1^3} \frac{\mbox{const.} k_1^{2\ell_1 + 2}}{|D(k_1)|^2} 
    \sim \frac{k_1^{2\ell_1 - 1}}{|D(k_1)|^2}  \;.
    \label{eq:sigma-reac-general}
\end{equation}
The elastic case with $f=1$ is treated separately in Section~\ref{sec:elastic}. 
The exact behavior of the cross sections will be dictated by that of $D(k_1)$.

We focus our attention on $D$, using $k\equiv k_1$ and $\ell \equiv \ell_1$ for clarity.
From its definition in Eq.(\ref{eq:def-det-D}) together with 
${\cal J}_{11}= A_{11}-iB_{11}$, we get
\[ D(k) = A_{11}-iB_{11} + {\sf j}_{11}\;, \]
where ${\sf j}_{11}=C_{11}^{-1}\sum_{n\neq1} (A_{1n}-iB_{1n})C_{1n}$ from Eq.(\ref{eq:def-det-J}).
Using the leading terms in Eq.(\ref{eq:A_1n+B_1n}) together with $C_{11}\sim C_{11}^{(0)} =\mbox{const.}$
and $C_{1n\neq 1} \sim C_{1n\neq 1}^{(0)} k^{\ell + 1}$ given in Eq.(\ref{eq:C_ij}).
we write
\begin{eqnarray} 
   {\sf j}_{11} & = & \frac{1}{C_{11}^{(0)}}  \sum_{n\neq1} 
     \left[ k^{-\ell- 1} \left( A_{1n}^{(0)} + k^2 A_{1n}^{(2)} + \dots \right) \right. \nonumber \\
     && \left. \hspace{.25in}
     -i k^{\ell} \left( B_{1n}^{(0)} + k^2 B_{1n}^{(2)} + \dots \right) \right] C_{1n}^{(0)} k^{\ell + 1} \;, \nonumber \\
     &=& \sum_{n\neq1} \frac{C_{1n}^{(0)}}{C_{11}^{(0)}}
     \left[ \left( A_{1n}^{(0)} + k^2 A_{1n}^{(2)} + \dots \right) \right. \nonumber \\
     && \left. \hspace{.25in}
     -i k^{2\ell + 1} \left( B_{1n}^{(0)} + k^2 B_{1n}^{(2)} + \dots \right) \right] \;,\nonumber \\
     & \equiv & {\sf j}_0 + {\sf j}_2 k^2 + \dots - i k^{2\ell + 1} \left( {\sf g}_0 + {\sf g}_2 k^2 + \dots \right) \;.
\end{eqnarray}
Here, the complex numbers ${\sf j}_i\equiv \sum_{n\neq1} A_{1n}^{(i)} C_{1n}^{(0)}/C_{11}^{(0)}$
and ${\sf g}_i\equiv \sum_{n\neq1} B_{1n}^{(i)} C_{1n}^{(0)}/C_{11}^{(0)}$
have small magnitudes. Together with Eq.(\ref{eq:A_11+B_11}),
we obtain
\begin{eqnarray}
   D(k) & = & \left[ (A_0 + {\sf j}_0) + (A_2 +  {\sf j}_2) k^2 + \dots \right] \nonumber \\
     && \hspace{-.15in}
   -i k^{2\ell + 1} \left[ (B_0 + {\sf g}_0) + (B_2 + {\sf g}_2) k^2 + \dots \right], 
   \label{eq:D(k)}
\end{eqnarray}
where we use the simpler notation  $A_i\equiv A_{11}^{(i)}$ and $B_i\equiv B_{11}^{(i)}$.
The exact form of $D(k)$ depends on the value of $\ell$, and  for this
reason, we consider $\ell =0$ and $\ell \geq 1$ separately.

\noindent $\bullet$ $\underline{\ell =0}$: in this case, we have
\begin{eqnarray*}
   D(k) & = & \left[ (A_0 + {\sf j}_0) + (A_2 +  {\sf j}_2) k^2 + \dots \right] \nonumber \\ && 
   -i k \left[ (B_0 + {\sf g}_0) + (B_2 + {\sf g}_2) k^2 + \dots \right] \;,  \\
   & = & D_0 + D_1 k + D_2 k^2 + D_3 k^3 + \dots
\end{eqnarray*}
where $D_0=A_0 + {\sf j}_0$, $D_1=-i (B_0 + {\sf g}_0)$, $D_2=(A_2 +  {\sf j}_2)$,
$D_3= -i(B_2 + {\sf g}_2)$, and so on. The expansion of $|D|^2$ takes the
form
\begin{equation}
  |D(k)|^2 \simeq \Delta_0 + \Delta_1 k + \Delta_2 k^2 + \dots \;, 
  \label{eq:D^2-expansion-l=0}
\end{equation}
with $\Delta_0 = |D_0|^2$, $\Delta_1 = D^*_0D_1+D_0 D^*_1$, $\Delta_2 = |D_1|^2+D^*_0D_2+D_0D^*_1$, etc.
The denominator $|D|^2$ will exhibit the Wigner or NTR scaling depending on the 
magnitude of $A_0$. If $A_0$ is dominant, then
$D_0$ is also sizable and $\Delta_0$ is the leading term in Eq.(\ref{eq:D^2-expansion-l=0}) 
for small $k$. However, if $A_0$ itself small, and since the magnitude of ${\sf j}_0$ 
is also small, there is a range of $k$ where $\Delta_0$ is not the dominant contribution, 
and since $\Delta_1$ is also proportional to $D_0$,
the next leading term is $\Delta_2\approx |D_1|^2$ (since both $D_0D^*_2$ and 
$D^*_0D_2$ must also be small). 
This condition gives the NTR scaling for a given range of $k$. 
To understand the Wigner and NTR regimes, $\Delta_1$ can be omitted in 
Eq.(\ref{eq:D^2-expansion-l=0}) since it plays a role only in the transition 
between the two regimes. Combining these results with Eq.(\ref{eq:sigma-reac-general})
gives (with $\ell =0$)
\begin{equation} 
   \sigma_{\ell =0}^{\rm inel.}\equiv \sigma^{\rm inel. (\ell=0)}_{f\leftarrow 1} 
   \sim \frac{k^{- 1}}{\Delta_0 + k^2 \Delta_2} \;.  
   \label{eq:NTR-CX-expansion}
\end{equation}
From Eqs.(\ref{eq:sigma_inel_CD}) and (\ref{eq:sigma-reac-general}), we note
that this $k$-scaling is the same for any exit channel $f\neq 1$, although each channel has
its specific magnitude.
The appearance of the NTR scaling depends of the relative strength of $\Delta_0$ and $\Delta_2$. From
Eq.(\ref{eq:NTR-CX-expansion}), we have for inelastic processes
\begin{equation}
   \sigma_{\ell =0}^{\mathrm{inel.}} \sim \left\{
   \begin{array}{lll}
        k^{-1} \;, & \mbox{Wigner: }  & k\ll\sqrt{|\Delta_0/\Delta_2|} ,\\
        k^{-3} \;, & \mbox{NTR: }      & k\gg\sqrt{|\Delta_0/\Delta_2|},
   \end{array}
   \right.
   \label{eq:NTR-reaction-expansion}
\end{equation}

\noindent $\bullet$ $\underline{\ell \neq 0}$: we first consider $\ell =1$. From Eq.(\ref{eq:D(k)}),
we have
\begin{eqnarray*}
   D(k) & = & \left[ (A_0 + {\sf j}_0) + (A_2 +  {\sf j}_2) k^2 + \dots \right] \nonumber \\ && 
   -i k^3 \left[ (B_0 + {\sf g}_0) + (B_2 + {\sf g}_2) k^2 + \dots \right] \;,  \\
   & = & D_0 + D_2 k^2 + \tilde{D}_3 k^3 + D_4 k^4 + \dots
\end{eqnarray*}
where $D_{0}=A_{0} + {\sf j}_{0}$ and  $D_2=(A_2 +  {\sf j}_2)$ as before,
$D_{4}=A_{4} + {\sf j}_{4}$, and $\tilde{D}_3=-i (B_0 + {\sf g}_0)$ (same as $D_1$ in the $\ell =0$ case), and so on. The expansion of $|D|^2$ becomes
\begin{equation}
  |D(k)|^2 \simeq \Delta_0 + \Delta_2 k^2 + \Delta_3 k^3 + \Delta_4 k^4 + \dots \;, 
  \label{eq:D^2-expansion-l=1}
\end{equation}
with $\Delta_0 = |D_0|^2$, $\Delta_2 = D^*_0D_2+D_0 D^*_2$, 
$\Delta_3 = D^*_0\tilde{D}_3+D_0 \tilde{D}^*_3$,
$\Delta_4 = |D_2|^2+D^*_0D_4+D_0D^*_4$, etc.
Again, when $A_0$ is dominant, $D_0$ is also sizable, and $\Delta_0$ is the leading term in 
Eq.(\ref{eq:D^2-expansion-l=1}) for small $k$, corresponding to the Wigner regime. If 
$A_0$ (and ${\sf j}_0$) is small, then $D_0$ is small, and
there is a range of $k$ for which $\Delta_0$, $\Delta_2$ and $\Delta_3$ are small compared to 
$\Delta_4\approx |D_2|^2$. As in the $\ell =0$ case, $\Delta_2$ and $\Delta_3$ play a role in the transition between the Wigner (with $\Delta_0$ dominant) and NTR (with $\Delta_4$ dominant) regimes, and can be omitted to describe the two regimes. We write
\begin{equation}
  |D(k)|^2 \approx \Delta_0 + \Delta_4 k^4 + \dots \;. 
  \label{eq:D^2-expansion-l=1-approx}
\end{equation}
Similarly, for $\ell=2$, Eq.(\ref{eq:D(k)}) gives
\begin{eqnarray*}
   D(k) & = & \left[ (A_0 + {\sf j}_0) + (A_2 +  {\sf j}_2) k^2 + \dots \right] \nonumber \\ && 
   -i k^5 \left[ (B_0 + {\sf g}_0) + (B_2 + {\sf g}_2) k^2 + \dots \right] \;,  \\
   & = & D_0 + D_2 k^2 + D_4 k^4 +  \tilde{D}_5 k^5 \dots
\end{eqnarray*}
where, $D_0$,  $D_2$, and $D_4$ are given above, and 
$\tilde{D}_5=-i (B_0 + {\sf g}_0)$ (same as $D_1$ in the $\ell =0$ case), and so on. 
The expansion of $|D|^2$ becomes
\begin{equation}
  |D(k)|^2 \simeq \Delta_0 + \Delta_2 k^2 + \Delta_4 k^4 + \dots \;, 
  \label{eq:D^2-expansion-l=2}
\end{equation}
with $\Delta_0$, $\Delta_2$, $\Delta_4$ are the same as for $\ell =1$. 
There is no $k^3$ term, which holds for $\ell >2$ in general. Again, 
for a sizable $A_0$, $\Delta_0$ is the leading term in Eq.(\ref{eq:D^2-expansion-l=2})
at small $k$, and for small $A_0$, there is a range of $k$ for which 
$\Delta_4\approx |D_2|^2$ is the leading term: $\Delta_2$ plays a role in the 
transition between the Wigner (with $\Delta_0$ dominant) and NTR 
(with $\Delta_4$ dominant) regimes, and is omitted. The same expression for $|D|^2$
can therefore be used for $\ell =1$ and $\ell\geq 2$, namely
\begin{equation}
  |D(k)|^2 \approx \Delta_0 + \Delta_4 k^4 + \dots \;. 
  \label{eq:D^2-expansion-l=2-approx}
\end{equation}
Combining this result with Eq.(\ref{eq:sigma-reac-general}), we
get
\begin{equation} 
   \sigma_{\ell\neq 0}^{\rm inel.}\equiv \sigma^{\rm inel. (\ell\neq 0)}_{f\leftarrow 1}  \sim \frac{k^{2\ell - 1}}{\Delta_0 + k^4 \Delta_4} \;. 
   \label{eq:NTR-CX-expansion-l}
\end{equation}
As in the $\ell=0$ case in Eq.(\ref{eq:NTR-CX-expansion}), this $k$-scaling is the same for 
any exit channel $f\neq 1$, each channel having its specific magnitude.
Eq.(\ref{eq:NTR-CX-expansion-l}) shows that the appearance of the NTR scaling depends 
of the relative strength of $\Delta_0$ and $\Delta_4$. 
\begin{equation}
   \sigma_{\ell\neq 0}^{\rm inel.} \sim \left\{
   \begin{array}{lll}
        k^{2\ell -1} \;, & \mbox{Wigner: }  &  
                          k\ll |\Delta_0/\Delta_4|^{1/4} ,\\
        k^{2\ell -5} \;, & \mbox{NTR: }      &  
                         k\gg  |\Delta_0/\Delta_4|^{1/4},
   \end{array}
   \right.
   \label{eq:NTR-reaction-expansion-l}
\end{equation}
We note that for $\ell =1$, the scaling leads to a $k^{-3}$ NTR
scaling for inelastic processes, as for $\ell=0$. This is illustrated 
for the benchmark system H$_2$+D in Fig.~\ref{fig:p-NTR}. 

In general, the NTR regime appears when $\Delta_0$ is small when compared to 
$\Delta_2$ (for $\ell=0$) or $\Delta_4$ (for $\ell\geq 1$). The transition between the 
Wigner and NTR regimes takes place  around $k=\sqrt{\Delta_0/\Delta_2}$ for $\ell=0$
or  around $k=\sqrt{\Delta_0/\Delta_4}$ for $\ell>0$. In \cite{Simbotin2015}, we
explored this transition in H$_2$+Cl for $\ell=0$ for the three resonances shown in 
Fig.~\ref{fig:alpha-beta+CX} by plotting the reaction probability $P=1-|S_{ii}|^2$,
a smooth function, also showing the effect of the reactivity (or background cross section
away from the resonance: see \cite{Simbotin2015} for more details).

\subsection{Simple model}

Since resonances usually become narrower with higher $\ell$, and occur at higher
scattering energies, the computational cost for benchmark systems containing H$_2$
quickly become prohibitive. Instead, we illustrate the effect of NTRs on cross sections
using a simpler model incorporating the key ingredients while allowing for easy
tuning of the resonances for each partial wave $\ell$.

\begin{figure}[ht]
\centerline{\includegraphics[clip,width=1.0\linewidth]{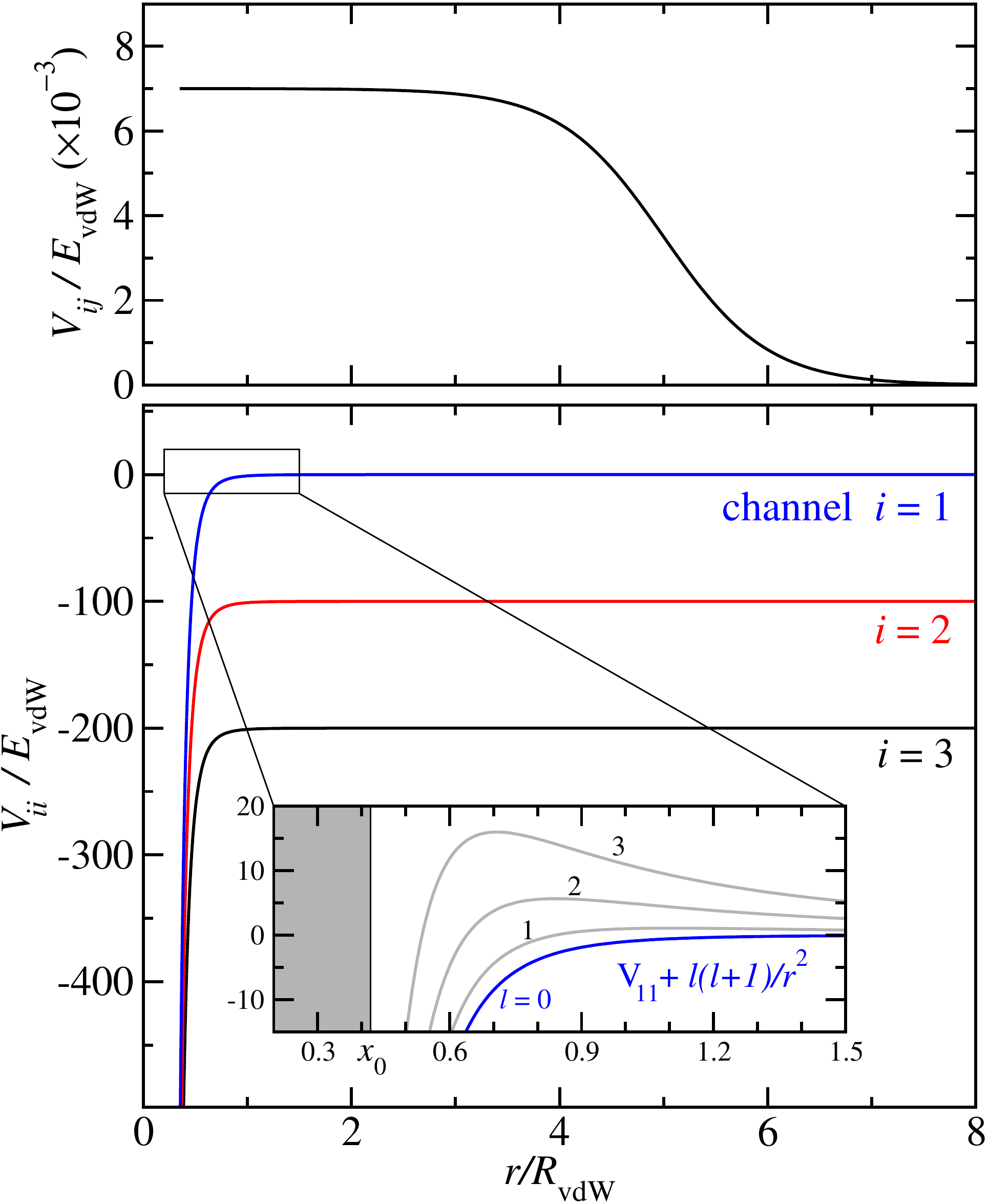}}
\caption{Upper panel:  coupling potential, see Eq.~(\ref{eq:couplings}), used in our
three-channel model.  Lower panel: diagonal potentials, see
Eq.~(\ref{eq:diagonals}).  The inset shows the effective potential in the
entrance channel for $\ell\leq 3$.}
\label{fig:vdiagonal}
\end{figure}

Fig.~\ref{fig:vdiagonal} depicts the model: it consists of three open channels with an attractive
$r^{-6}$ potential tail and a short-range hard wall. The position $r_0$ of the hard 
wall is shifted slightly
for each partial wave as to bring a resonance in the entrance channel $i=1$. The diagonal
potentials $V_{ii}$ are identical for each channel $i$, each with their own threshold $E_i$.
To simplify notations, we use van der Waals units $R_{\rm vdW}$ for the length and
$E_{\rm vdW}=\hbar^2/2\mu R_{\rm vdW}^2$ for energy, where $\mu$ is the reduced
mass of the scattering partners. For an attractive power-law tail 
$V(r)\sim -C_\alpha r^{-\alpha}$, the van der Waals length scale 
is $R_{\rm vdW} = (2\mu C_\alpha/\hbar^2 )^{\frac{1}{\alpha-2}}$. The off-diagonal couplings
$V_{ij}$ are taken to be identical and short-range. Defining $x\equiv r/R_{\rm vdW}$, the
diagonal and off-diagonal potentials have the form
\begin{eqnarray}
   V_{nn}(r)  & = & \left\{ \begin{array}{ll} \displaystyle
                              +\infty & \mbox{, for } r\leq r_0 \;,   \\
    \displaystyle -\frac{E_{\rm vdW}}{x^6 } + E_i & \mbox{, for } r>r_0 
                               \end{array} \right. \;,  \label{eq:diagonals}\\
   V_{ij} = & = & \frac{0.007 E_{\rm vdW}}{1 + \exp[2(x- 5)]} \;. \label{eq:couplings}
\end{eqnarray}
The energy threshold for each channel $i$ and the values of $r_0$ bringing a near threshold resonance in the entrance channel for a given partial wave $\ell$ are respectively
\begin{equation}
   \frac{E_i}{E_{\rm vdW}} = \left\{ \begin{array}{ll}
                 0 & \mbox{ , for channel 1} \\
            -100 & \mbox{ , for channel 2} \\
             -200& \mbox{ , for channel 2} \end{array} \right. \;,
\end{equation}
and
\begin{equation}             
   \frac{r_0}{R_{\rm vdW}} = \left\{ \begin{array}{ll}           
     0.42402677 & \mbox{ , for $\ell =0$} \\ 
     0.37845091 & \mbox{ , for $\ell =1$} \\
     0.34646173& \mbox{ , for $\ell =2$} \\
     0.32219300& \mbox{ , for $\ell =3$}  \end{array}\right. \;.
\end{equation}
For simplicity sake, only the entrance channel $i=1$ contains the centrifugal term $\ell(\ell+1)/x^2$.

\begin{figure}[ht]
\centerline{\includegraphics[clip,width=1.0\linewidth]{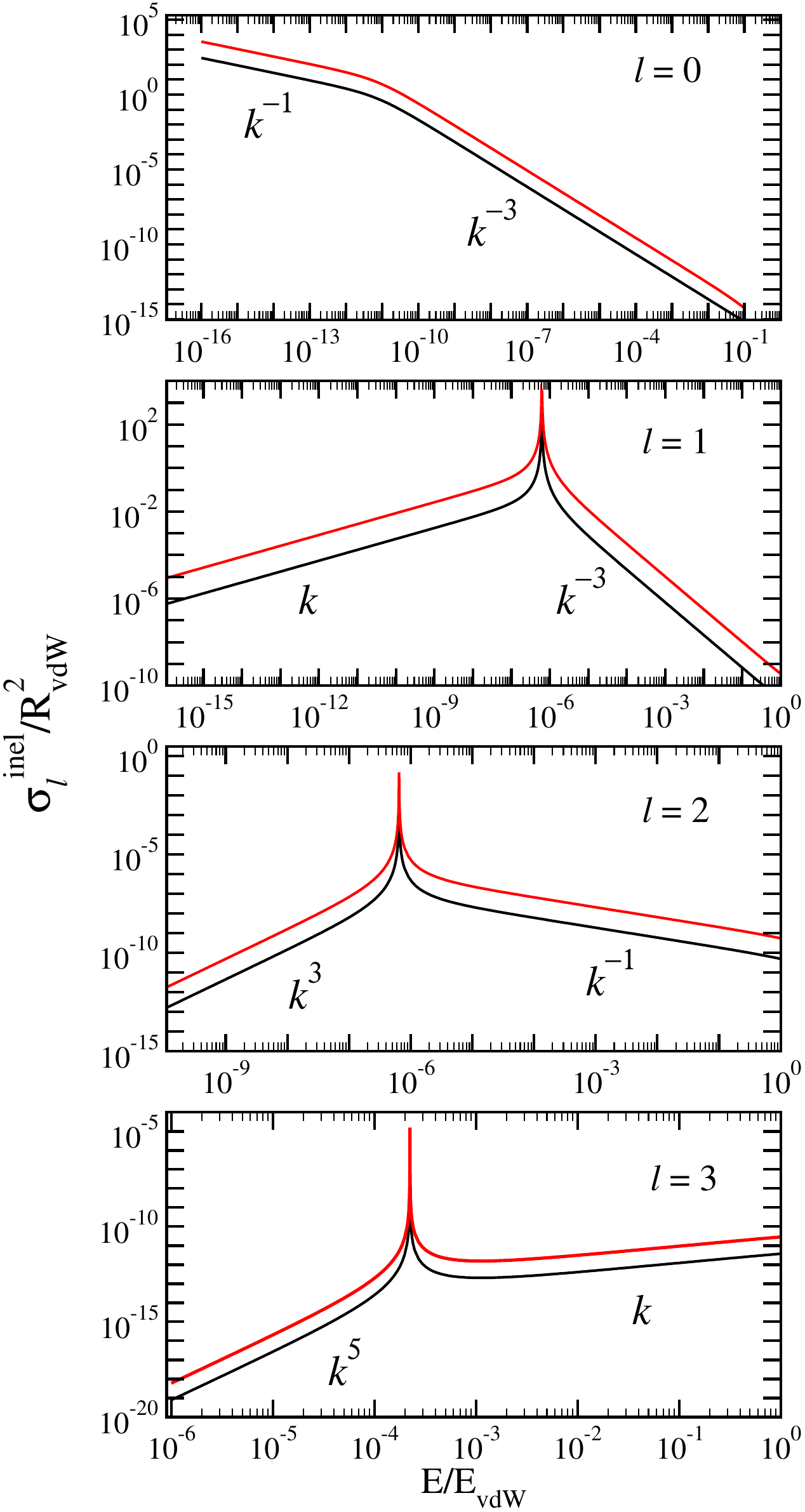}}
\caption{Simple model: individual partial inelastic cross sections 
              $\sigma^{\rm inel. (\ell)}_{f\leftarrow 1}$
              for $\ell =0, \dots , 3$ as a function of the scattering energy $E$. Both $E$ and
              $\sigma$ are given in scaled van der Waals units $E_{\rm vdW}$ and $R_{\rm vdW}$,
              respectively, with different range for the cross section for each $\ell$. 
              Red and black curves correspond to the different final channels $i=1\rightarrow f=2$
              and $i=1\rightarrow f=3$, respectively.
}
\label{fig:inelastic}
\end{figure}

We compute the partial cross sections $\sigma^{\rm inel. (\ell)}_{f\leftarrow 1}$ for
each $\ell$ and for the final channel being $2$ or $3$. The results are shown in
Fig.~\ref{fig:inelastic}, where the axes of each panel have different ranges due to the
changing resonance width and position. They demonstrate that for both final channels,
the inelastic partial cross section follows the $k$-scaling given by 
Eq.(\ref{eq:NTR-reaction-expansion-l}) for $\ell=0$ and Eq.(\ref{eq:NTR-reaction-expansion-l})  
for $\ell \geq 1$. More specifically, it verifies that the NTR regime scaling multiplies
the Wigner regime by $k^{-2}$ for $\ell=0$, and by $k^{-4}$ otherwise. Since the 
resonance is in the entrance channel, the partial inelastic cross section into the two
remaining final channels are besically identical within an overall constant.

\subsection{Elastic processes}
\label{sec:elastic}

In the elastic case, the range of the interaction potential may play an important role
in the $k$-scaling of the cross section. We therefore consider short-range and
long-range (actually power-law type) interactions separately.

\subsubsection{Short-range}

\begin{figure}[ht]
\centerline{\includegraphics[clip,width=1.0\linewidth]{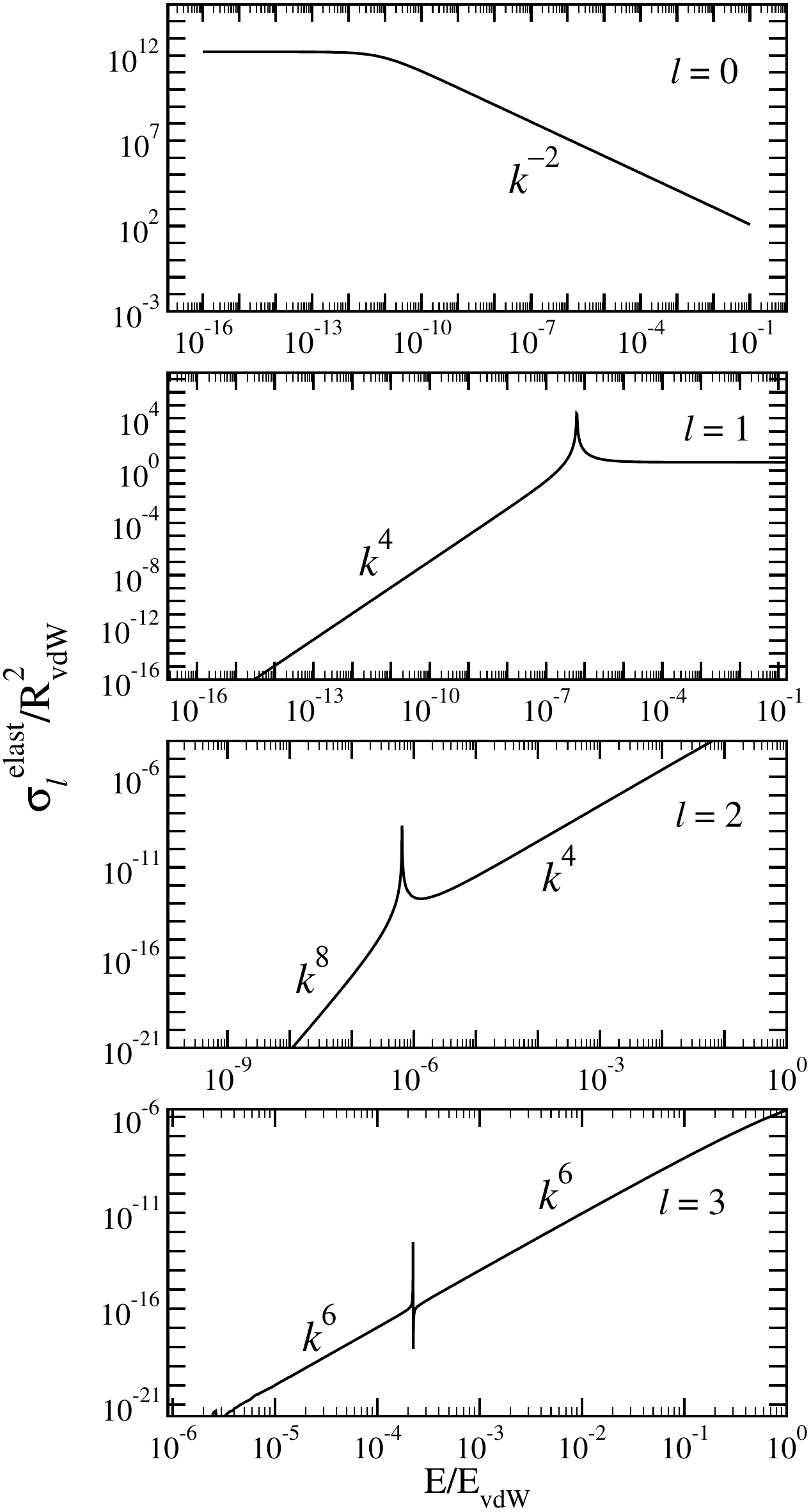}}
\caption{Same as Fig.~4 for the elastic cross section;  see text for discussion.}
\label{fig:elastic}
\end{figure}

The previous treatment applies to this case ({\it e.g.}, for interaction with long-range 
exponential tail like the Morse-type potential). For $f=1$, and replacing $k_f=k_1\equiv k$
in Eq.(\ref{eq:sigma_inel_CD}), and using the results of Eq.(\ref{eq:C_ij}), we have
$\sum_j B_{1j} C_{1j} \sim k^{2\ell + 1}$,
so that
\begin{equation} 
    \sigma^{\rm elas}_{1\leftarrow 1} \sim  \frac{k}{k^3} \frac{\mbox{const.} k^{4\ell+ 2}}{|D(k)|^2} 
    \sim \frac{k^{4\ell}}{|D(k)|^2} \;.
    \label{eq:sigma-elas-general}
\end{equation}
The previous results for $|D(k)|^2$ apply here as well, and we obtain
for $\ell =0$
\begin{equation} 
   \sigma_{\ell =0}^{\rm elas} \sim \frac{k^{0}}{\Delta_0 + k^2 \Delta_2} \;,
   \label{eq:NTR-CX-elas-expansion}
\end{equation}
leading to
\begin{equation}
   \sigma_{\ell =0}^{\mathrm{elast}} \sim \left\{
   \begin{array}{lll}
      k^0   & \mbox{, Wigner: }  & k\ll\sqrt{|\Delta_0/\Delta_2|} ,\\
      k^{-2}  & \mbox{, NTR: }      & k\gg\sqrt{|\Delta_0/\Delta_2|}.
   \end{array}
   \right.
\end{equation}
For $\ell \geq 1$, we get
\begin{equation} 
   \sigma_{\ell \neq 0}^{\rm elas} \sim \frac{k^{4\ell}}{\Delta_0 + k^4 \Delta_4} \;,
   \label{eq:NTR-CX-expansion-l-elas}
\end{equation}
leading to
\begin{equation}
   \sigma_{\ell\neq 0}^{\mathrm{elast}} \sim \left\{
   \begin{array}{lll}
      k^{4\ell} & \mbox{, Wigner: } &  k\ll |\Delta_0/\Delta_4|^{1/4} ,\\
      k^{4\ell-4} & \mbox{, NTR: }    &  k\gg  |\Delta_0/\Delta_4|^{1/4}.
   \end{array}
   \right.
   \label{eq:elas-high-l}
\end{equation}
We note that the NTR regime scales as $k^0$ for $\ell =1$.
Fig.~\ref{fig:elastic} shows the elastic cross section for the model of the previous
section, {\it i.e.} a long-range tail of the form $-C_6/r^6$. For both $\ell=0$ and 1,
we observe the expected scalings for the Wigner and NTR regimes. However, the
$k$-scaling for higher $\ell$ values does not seem to follow Eq.(\ref{eq:elas-high-l});
for $\ell=3$, below and above the resonance, the scaling follows $k^6$ instead of
the expected $k^{12}$ (Wigner) and $k^{8}$ (NTR) scalings. This is due to the
power-law long-range tail of the interaction potential. As we will see below, even for
$\ell =2$, the $k$-scaling shown in Fig.~\ref{fig:elastic}, though seemingly agreeing with
the short-range scaling $k^8$ (Wigner) and $k^4$ (NTR) given by Eq.(\ref{eq:elas-high-l}),
it is actually not following the appropriate Wigner scaling regime. To understand these
details, we consider the effect of the power-law tail on the elastic cross section.

\subsubsection{Power-law tail}

While the threshold behavior (including both NTR and Wigner regimes) of
the inelastic cross sections is unaffected by the long-range nature of
the diagonal potential in the entrance channel, the elastic cross
section at low energy can be altered significantly by the long-range
tail of $V_{11}(r)$.  This can be understood in terms of the
single-channel Jost function, $\mathcal{J}=A-iB$, corresponding to the entrance
channel ($n=1$).  Using $S=\mathcal{J}^*\mathcal{J}^{-1}$, the partial 
single-channel elastic cross section (for a given $\ell$) is
simply 
\begin{equation}
   \sigma^{\rm elas}_\ell = \frac{4\pi}{k^2} |1-S_\ell (k)|^2 
   = \frac{4\pi}{k^2} \frac{B^2(k)}{A^2(k) + B^2(k)} \;,
   \label{eq:elas-single-ch}
\end{equation}
where we omit the subscript $\ell$ for $A_\ell$ and $B_\ell$ for clarity.
According to Willner and Gianturco
\cite{gianturco}, the $k$-dependence of the single-channel Jost
function for a potential which behaves asymptotically
($r\rightarrow\infty$) as an inverse power, $V(r)\approx r^{-\alpha}$,
takes the form
\begin{equation} \left. \begin{array}{lll}
A(k) & = & \tilde A(k) L^{AA}(k) + \tilde B(k) L^{AB}(k) \;, \\
B(k) & = & \tilde B(k) L^{BB}(k) + \tilde A(k) L^{BA}(k) \;,
\end{array} \right\}
\end{equation}
where $\tilde A(k)$ and $\tilde B(k)$ are analytic functions,
\begin{equation} \left. \begin{array}{lll}
\tilde A(k) & = & A_0 + A_2 k^2 +\cdots \;, \\
\tilde B(k) & = & k^{2\ell+1}(B_0+B_2k^2 +\cdots) \;, 
\end{array} \right\}
\end{equation}
while the functions $L(k)$ contain the effect of the long-range tail,
and can be expanded as power series (possibly including log-terms)
\cite{gianturco},
\begin{equation} \left. \begin{array}{ll}
L^{AA}(k) & \!\! = 1  \! + \! a(C_\alpha k^{\alpha-2}) \! + \!  a'(C_\alpha k^{\alpha-2})^2 \! + \!  \cdots \;, \\
L^{BB}(k) & \!\! = 1 \! + \!  b(C_\alpha k^{\alpha-2}) \! + \!  b'(C_\alpha k^{\alpha-2})^2 \! + \!  \cdots \;, \\
L^{AB}(k) & \!\! =  c(C_\alpha k^{\alpha-2}) \! + \!  c'(C_\alpha k^{\alpha-2})^2 \! + \!  \cdots \;, \\
L^{BA}(k) & \!\! = d(C_\alpha k^{\alpha-2}) \! + \!  d'(C_\alpha k^{\alpha-2})^2 \! + \!  \cdots \;,
\end{array} \!\! \right\}
\end{equation}
where $a,\ a'\,\ b,\ b',\ c,\ c',\ d,\ d',\ldots$, are constants.

In general, one must keep both ``normal'' and ``mixed/cross'' terms when
truncating the low-k expansions for $A$ and $B$, giving
\begin{equation} \left. \begin{array}{ll}
A(k) & \!\! \approx \!  A_0 \! +\! A_2 k^2 \! +\!  B_0 c C_\alpha k^{2\ell+\alpha-1} + \cdots \;, \\
B(k) & \!\! \approx  \! (A_0 \! +\! A_2 k^2)dC_\alpha k^{\alpha-2} \! +\!  B_0 k^{2\ell+1}
+ \cdots .
\end{array} \!\! \right\}
\label{eq:A-B-alpha}
\end{equation}
Thus, unlike the short-range case, the $k$-dependence of the Jost
function for a long-range potential is considerably more complex.  In
particular, for $\ell \geq \frac{1}{2}(\alpha-3)$, the dominant term
for $B(k)$ at low-$k$ will no longer be $B_0k^{2\ell+1}$, but
$A_0dC_\alpha k^{\alpha-2}$ instead.  Consequently, when $A_0$ is
vanishingly small (NTR case), both $A(k)$ and $B(k)$ can lose their
dominant term simultaneously. 
. This is particularly important for the elastic cross section 
in Eq.({\ref{eq:elas-single-ch}), since it contains $B(k)$ in the numerator.

\subsubsection{$\alpha =6$}

Let us explore the specific case $\alpha=6$ corresponding to our model
and most interactions for neutral ground state scattering partners (without
permanent dipole or quadrupole moments).
In that case, the critical (transition) value for the angular momentum is
$\ell_*=\frac{\alpha-3}{2}=\frac{3}{2}$. Thus, for $s$-wave and $p$-wave,
the leading $k$ powers in $A$ and $B$ are the same as the short-range
case, and so is the low-$k$ behavior of the elastic cross section, while for 
$d$-wave and higher ($\ell\geq\frac{3}2$) we expect new types of behavior.

\subsection*{Partial wave $\ell=2$}
According to Eq.(\ref{eq:A-B-alpha}), with $\alpha=6$ and $\ell=2$, we have
\begin{eqnarray*}
   A(k) & \approx &  A_0 + A_2 k^2 + B_0 c C_6 k^5 + \cdots \;, \\
   B(k) & \approx & (A_0 + A_2 k^2)dC_6 k^{4} \! +\!  B_0 k^{5} + \cdots \;.
\end{eqnarray*}
In the absence of NTR, $A_0$ is sizable, and $A(k)\sim A_0$ while 
$B(k)\sim A_0dC_6 k^{4} \propto k^4$, so that the Wigner regime behavior 
of the elastic cross section should be
\begin{equation}
   \sigma^{\rm elas}_{\ell=2} 
   = \frac{4\pi}{k^2} \frac{B^2}{A^2+ B^2} \propto k^6 \;, 
   \mbox{ bare/true Wigner}\; .
\end{equation}
However, when a shape resonance is very close to the threshold, $A_0$
becomes vanishingly small, and the Wigner regime practically
disappears into the very-very-deep ultracold. In Fig.~\ref{fig:elastic}, it
would be visible at much lower energies (not shown).

Indeed, for $\ell=2$, $B(k)\approx A_0dC_6 k^4 + B_0 k^5$,
and the competition between the leading order term ($A_0dC_6k^4$) and
the next order term ($B_0 k^5$) leads to a transition at around 
$k_B\sim|A_0dC_6/B_0|$. However, the denominator $A^2+B^2$ is dominated by 
$A(k)\approx A_0+A_2 k^2$ at small $k$, giving to a transition between
the leading term ($A_0$) and the next order term ($A_2 k^2$) at 
$k_A\sim|A_0/A_2|^{1/2}$. For the NTR condition, $A_0$ becomes small,
and although  both $k_B$ and $k_A$ vanish with $A_0$, $k_B$ vanishes 
much faster than $k_A$, and a new (intermediate) regime appears.  This new regime
($k_B<k<k_A$) can be regarded as the (effective) Wigner regime,
because the (bare/true) Wigner regime itself ($k<k_B$) is lost in the deep
ultracold.

Within the new (effective Wigner) regime, the $A_0$ term is negligible
in the numerator so that $B^2(k)\sim B_0^2 k^{10}$, but it is still dominant in
the denominator $A^2(k)$, and we have:
\begin{equation}
    \sigma^{\rm elas}_{\ell=2}  \sim \frac{4\pi}{k^2} \frac{B_0^2k^{10}}{A_0^2} \sim k^8,
     \mbox{ effective Wigner regime.}
\end{equation}
As mentioned above, the $\ell =2$ $k$-scaling in Fig.~\ref{fig:elastic} shows the 
effective Wigner regime $k^8$ scaling, the bare/true Wigner $k^6$ regime appearing 
at much lower energies (not shown).

\subsection*{Partial waves $\ell=3$ and higher}

For $\ell\geq 3$, the leading orders for $A$ and $B$ 
according to Eq.(\ref{eq:A-B-alpha}) are
$A(k)\approx A_0 + A_2 k^2$, and
 $B(k)\approx (A_0 + A_2 k^2)dC_6 k^4$.  The
$B_0$ terms can be neglected because they are of higher order ($k^{11}$ in $A$ and 
$k^7$ in $B$ for $\ell=3$).  Thus, except in the immediate vicinity of the very
narrow shape resonance, we have:
\begin{equation}
\sigma^{\rm elas}_{\ell\geq 3} \sim \frac{4\pi}{k^2} \frac{\big[(A_0+A_2 k^2)dC_6 k^4\big]^2}{(A_0+A_2
  k^2)^2}
 \propto k^6.
\end{equation}
Hence, there is only one power law for both Wigner and NTR regimes
(with a narrow spike/resonance in the middle), as depicted in Fig.~\ref{fig:elastic}.

\section{conclusion}

In this paper, we investigated the effect of near threshold resonances (NTRs) on both
the elastic and inelastic cross sections for given partial waves $\ell$ at low scattering 
energies. In particular, we considered benchmark reactions involving molecular hydrogen,
H$_2$+Cl (for $s$-wave) and H$_2$+D (for $p$-wave). The later possesses resonant
features that are reachable experimentally. For higher partial waves, we used a three open
channel model incorporating the key ingredients relevant to NTRs. The interaction
potentials in all those cases have a $r^{-6}$ long-range tail. We numerically found
that the inelastic cross sections follows two $k$-scaling laws, namely 
$\sigma^{\rm inel.}_{\ell =0}\sim k^{-1}$ (Wigner) and $k^{-3}$ (NTR), and
$\sigma^{\rm inel.}_{\ell \neq 0}\sim k^{2\ell -1}$ (Wigner) and $k^{2\ell -5}$ (NTR).
These scalings follow those obtained by analyzing the analytical behavior of the
inelastic cross section for short-range interactions, based on Jost functions. This is
to be expected, since inelastic scattering processes are due to short-range couplings
overtaking the long-range tail of the diagonal term of the interaction matrix.

The case of elastic scattering is slightly different.  The results for short-range interactions 
were found to be
$\sigma^{\rm elas}_{\ell =0}\sim k^{0}$ (Wigner) and $k^{-2}$ (NTR), and
$\sigma^{\rm elas.}_{\ell \neq 0}\sim k^{4\ell}$ (Wigner) and $k^{4\ell -4}$ (NTR).
However, the long-range tail of the interaction affects some partial waves.
In the $r^{-6}$ case considered here, we found that $\ell=0$ and 1 follow the short-range
results, but for $\ell\geq 2$, the power-law tail modifies those scalings. For
$\ell=2$, the true Wigner regime scales like $k^6$ (instead of the short-range $k^8$),
while the NTR regime scales like the expected $k^4$; however, we witnessed the
appearance of an effective Wigner regime scaling as $k^8$ between those two
regimes. For $\ell >2$, the cross section follows a $k^6$ scaling which does not agree
with either the Wigner or the NTR scalings for short-range potentials. These
results hint at a vanishingly relevant effect of a resonance on the elastic cross section 
with increasing $\ell$; beside a sharp and narrow feature at resonance, the $k$-scaling
is ``monotonic".

Understanding the effect of near threshold resonances on scattering processes
is important to help understanding but also predict the behavior of ultracold
systems. The different $k$-scaling of elastic and inelastic cross sections can help
guiding experimental efforts at cooling ultracold molecular samples. 

Again, understanding the behavior of ultracold samples, atomic or molecular, requires understanding
the role played by resonances, such as NTRs. They dictate the behavior of ultracold systems, and
can be used to manipulate and control processes in these systems.

\section*{Acknowledgments}
This work was partially supported by the MURI US Army Research Office
Grant No. W911NF-14-1-0378 (IS) and by the US Army Research Office,
Chemistry Division, Grant No. W911NF-13-1-0213 (DS, RC).

\bibliography{BOOK-RESONANCE_bib,ntr}

\end{document}